\documentstyle{mn}
\input psfig.sty

\begin{document}
\title[Small-Scale Clustering]
{On the Small-Scale Clustering of the ${\rm Ly}\alpha$ Forest Clouds}

\author[ M. M. \'Cirkovi\'c \& K. M. Lanzetta]
{Milan M. \'Cirkovi\'c$^{1,2}$ \& Kenneth M. Lanzetta$^2$
\\
 $^{1}${Astronomical Observatory Belgrade, Volgina 7,
11000 Belgrade, SERBIA} \\
{\tt mcirkovic@aob.aob.bg.ac.yu}\\
$^2$ Astronomy Program, Department of Physics and
 Astronomy \\
SUNY at Stony Brook, Stony Brook, NY 11794--3800,
U.S.A.
}

\date{Accepted .
      Received ;
      in original form }
\pubyear{2000}

\maketitle

\label{firstpage}
\begin{abstract}
Recent measurements of the autocorrelation function of the Ly$\alpha$ clouds are analyzed from the point of view of a simple model with strong clustering on the small scales. It is shown that this toy
model reproduces fairly well the important linear relation between amplitude of the absorber autocorrelation function and neutral
hydrogen column density. In addition, it predicts a correct
evolutionary trend of correlation amplitudes. Some possible
ramifications of these results are discussed.
\end{abstract}

\begin{keywords}
{quasars:  absorption lines---intergalactic medium---galaxies: haloes---galaxies: evolution}
\end{keywords}

\section{INTRODUCTION}
The problem of clustering of the Ly$\alpha$ forest absorbers
has been discussed in
the course of the last two decades by many authors (Sargent et
al.~1980; Dekel 1982; Salmon \& Hogan 1986; Webb 1987; Ostriker,
Bajtlik \& Duncan 1988; Crotts 1989; Heisler, Hogan \& White 1989;
Liu \& Jones 1990; Webb \& Barcons 1991; Barcons \& Webb 1990, 1991; Fang 1991; Mo et al.~1992; Fardal \& Shull 1993; Srianand \& Khare 1994; Chernomordik 1995; Elowitz, Green \& Impey 1995; Meiksin \& Bouchet 1995; Carbone \& Savaglio 1996; Srianand 1996; Ulmer 1996; Fern\'andez-Soto et al.~1996; Lanzetta, Webb \& Barcons 1996; Pando \& Fang 1996;
Cen \& Simcoe 1997; Rauch 1998).
The conclusions of the entire effort are still controversial,
since the original paradigm that ${\rm Ly}\alpha$ clouds do not
show no clustering at all (Sargent et al.~1980; Barcons \& Webb
1990), was somewhat undermined  by findings of weak clustering of intermediate-redshift ${\rm Ly}\alpha$ clouds (Webb 1987; Barcons
\& Webb 1991; Webb \& Barcons 1991; Chernomordik 1995),
and seriously questioned by the work on associated C IV absorption
(Cowie et al.\ 1995; Fern\'andez-Soto et al.~1996; see also Songaila \&
Cowie 1996). In an
interesting work, Crotts (1989) has investigated  correlations
in the real space across the sky among systems in multiple lines-of-sight,
and not only detected small-scale clustering, but also established the increase of clustering amplitudes with increasing column density, the conclusion which we shall quantify below. Difficulties and limitations
inherent in any attempt to use the two-point correlation function to
deduce the properties of the ${\rm Ly}\alpha$
forest are summarized in Rauch et al.~(1992) and Fern\'{a}ndez-Soto et al.~(1996). Theoretical analysis of Dekel (1982) was very
interesting in this respect, since it weakened the dominant paradigm
of Sargent et al.~(1980)---which, as we have seen, dictated much
of the future development of absorption-line studies---where it
looked strongest: in the cosmological part of the
argument. As Dekel writes: "My aim is to point out a cosmological
scenario in which galaxies are clustered only weakly at $z > 1.7$, so that the LACs [Ly$\alpha$ clouds] may cluster just like galaxies. Here
both the isothermal and the truncated adiabatic components of the
density perturbations play a role in the formation of structure in
the universe." It is also quite interesting to note that Dekel (1982) was
the first to suggest usage of C IV lines to measure the degree of
clustering of  absorption systems, idea which was fully realized
only 14 years later by Fern\'{a}ndez-Soto et al.~(1996). It
should certainly be mentioned that low spectral resolution of most
of existing measurements
(e.g.~$250 - 300$ km s$^{-1}$ for the {\it HST\/} Key Project;
see Jannuzi 1997; Jannuzi et al.\ 1998) makes investigations on small scales exceedingly difficult. And it is exactly these scales which are, because of their discriminative power, the most interesting from
our point of view.

Webb (1987) was the first to point out the presence of weak clustering at small velocity scales, based on the Voigt profile fitting in the
high-redshift data, which was confirmed by other investigations (e.g.~Muecket \& Mueller 1987). As emphasized by Fern\'andez-Soto
et al.~(1996), it is very difficult to directly detect
clustering of the Ly$\alpha$ lines because of the short redshift path length in any individual QSO spectrum, and line blending. If a
significant fraction, or most of the observed Ly$\alpha$
absorption lines are in fact blends of very narrow components,
detected amplitudes of the autocorrelation function would be
significantly underestimated (see also a discussion in Rauch
et al.~1992).

On the basis of these early findings, Ostriker
et al.~(1988) first proposed gravitationally induced clustering
as one of the possible explanations for excess of pairs of absorbers at small (and in their sample, intermediate\footnote{which has not been confirmed afterwards; see the discussion in Chernomordik (1995).})
velocity splittings.  However, in a recent important study, Cristiani et al.~(1997), performed the most comprehensive analysis of the clustering of ${\rm Ly}\alpha$ clouds in a sample of about 1600 absorbers along 15 lines of sight, and concluded that the small-scale clustering of ${\rm Ly}\alpha$ absorbers (a) is real, and (b) can be understood in terms of gravitationally-induced clustering, in the manner of Ostriker et
al.~(1988). Parenthetically, the existence of structure in the Ly$\alpha$ forest was confirmed by independent methods aimed at detection of the deviation of spatial distribution of absorbers from a uniform random one; thus Fang (1991) showed that Ly$\alpha$ forest deviate from a uniform distribution at 3$\sigma$ significant level. This is, of course, still weaker from the non-uniformity seen among the known galactic population, but very different from the picture of uniform, diffuse intergalactic population envisaged by Sargent et al.~(1980).

Another recent work of great importance for the development of our ideas on the spatial distribution of Ly$\alpha$ clouds is that of Ulmer (1996), who investigated a sample of low-$z$ Ly$\alpha$ lines recently obtained with the {\it HST\/} Key Project (Bahcall et al.\ 1996). Results of that work are particularly significant, since they demonstrate the existence
of strong clustering of Ly$\alpha$ lines at velocity separations at which high-$z$ lines seem completely unclustered, and which represent an intermediate regime between the small-scale ($\Delta v \leq 200$ km s$^{-1}$) and large scale clustering. In physical terms, we can
hypothesize that the small-scale regime can be plausibly explained as characteristic of the {\it intragalactic\/} motions in typical $L \sim L_\ast$ galaxies, being
on the same order as velocity dispersion in the known galactic
subsystems (Binney \& Tremaine 1987). On the other hand, large-scale clustering is
generally believed to trace large-scale structure (i.e.~structures with velocity dispersions similar to that in rich galaxy clusters and higher).
Ulmer (1996) has not obtained any information on the small velocity splittings, since his method explicitly rejects velocity splittings with
$\Delta v < 250$ km s$^{-1}$. Still, its implications are important because of the very strong signal found for $250 \leq \Delta v \leq 500$ km s$^{-1}$, which, if extrapolated into $\Delta v< 250$ km s$^{-1}$ agrees well with results of Cristiani et al.~(1997) and those discussed in further text. Similar excess of absorber pairs for $\Delta v \sim 200$ km s$^{-1}$ was found by Srianand \& Khare (1994) at the $\sim 4\sigma$ level. The redshift
evolution of clustering is also correctly emphasized in Ulmer (1996), who inferred a substantial increase in the degree of clustering of the
Ly$\alpha$ forest. As we shall see, there is evidence that the same
trend is real over most of the history of the universe.

In the rest of this paper, we shall attempt to show that the results
of Ulmer (1996) and Cristiani et al.~(1997) for the amplitude of TPCF are consistent with simple model characterized by constant small-scale clustering. Specifically, we shall show that (i) the linear relation between column density of clouds and their autocorrelation amplitude, and (ii) the general evolutionary trend of decrease in clustering with
increasing redshift, are successfully explained in such a toy model.
Following two sections are, therefore, devoted to these two
important issues. Although not specifically endorsing such a simplistic approach, it
does make the complex explanations for the observed
TPCF properties, involving biasing for the structure formation and gravitationally induced correlations, unnecessary.
In contradistinction, models in which the Ly$\alpha$ forest is
locally decoupled from the Hubble flow and physically associated with collapsed structures (e.g.~galaxies) predict, in general, exactly such a behavior.

\section{A SIMPLE MODEL}

Let the Ly$\alpha$ cloud distribution function (neglecting the Doppler parameter dependence) be written in the standard form as (e.g. Lu et al. 1996):
\begin{equation}
\label{jedan}
F(N,z) \equiv \frac{\partial n}{\partial z\; \partial N_{{\rm H\ I}}}=
A_0 (1+z)^\gamma N_{{\rm H\ I}}^{-\beta}.
\end{equation}
Constants in equation (\ref{jedan}) were measured by various authors (Hu et
al.\ 1995; Lu et al.\ 1996; Kim et al.\ 1997). For our purpose, it is enough to take approximate values $\gamma = 2.75$, and $\beta = 1.55$
for the indices of redshift and column density distribution respectively. One of the ways to simplify relation (\ref{jedan}) is to consider only column
density dependence in a sufficiently large sample of absorbing lines.
This column density distribution function we shall denote by $f(N)$,
and its standard functional form as
\begin{equation}
\label{dva}
f(N)=BN^{-\beta}.
\end{equation}
Normalization for $f(N)$ is given as $B=9.2 \times 10^8$ (Lu et
al.~1996).

Let us consider a simple model in which absorbers along the line of
sight are clustered around given points along the line of sight with
small-scale clustering
described by $\phi(v)$ in the form of the step function, such that
\begin{equation}
\label{mmm}
\phi (v) = \left\{ \begin{array} {r@{\quad:\quad}l} \phi(v) = \phi = {\rm const.} & v \leq \sigma_{\rm max} \\
0 & v > \sigma_{\rm max}
\end{array} \right.
\end{equation}
Here, $\sigma_{\rm max}$ is the maximum {\bf total} velocity dispersion characteristic for ${\rm Ly}\alpha$ absorption systems, i.e.~both intragalactic and intergalactic, although at this stage its physical
origin is not crucial. Following Fern\'andez-Soto et al.\ (1996), we
shall take a fiducial value $\sigma_{\rm max}= 150$ km s$^{-1}$
(see also Crotts 1989; Mo et al.\ 1992).

The TPCF is, in general, defined by the probability
\begin{equation}
\label{tri}
dP = (1+\xi) n_a dv.
\end{equation}
For our simple model of clouds concentrated around given points along the line of sight the differential probability of finding another cloud at velocity separation $dv$ is simply
\begin{equation}
\label{cetiri}
dP=\phi(v)dv + n_a dv.
\end{equation}
In these relations, $n_a$ is the average absorber density along the entire line of sight given as
\begin{equation}
\label{pet}
n_a(N_{\rm min}, z)= \int\limits_{N_{\rm min}}^{\infty} \! \!
\int\limits_0^z \! \! F(N, z) \; dN \,dz.
\end{equation}
Consistency requires that probabilities in eqs.~(\ref{tri}) and (\ref{cetiri}) are equal. It immediately follows that
\begin{equation}
\label{sest}
\xi = \frac{\phi(v)}{n_a}=\frac{\phi(v)}{\int_{N_{\rm min}}^\infty \! f(N)dN}= \frac{\beta-1}{B}\phi(v)N_{\rm min}^{\beta-1},
\end{equation}
which gives the amplitude of TPCF as a function of threshold column density $N_{\rm min}$. Assumption here is that the column density distribution function stays the same at all redshifts (i.e.~along the entire line of sight), enabling us to suppress the epoch dependence in (\ref{sest}).
Notice that in this model the quantity $\phi(v)$ is determined by the extrapolated unity column density correlation through relation
\begin{equation}
\label{sedam}
\log \xi(\log N = 0)= \log \phi(v) + \log \frac{\beta-1}{B}.
\end{equation}
Note that up to this point it does not matter whether $\phi (v)$ is
constant within some velocity range, as we supposed in eq.~(\ref{mmm}),
or not. From the mathematical point of view, this hypothesis remains
unnecessary; however, in order to establish firm contact with
correlation observations of necessarily very limited
velocity resolution, we
shall henceforth explicitly assume $\phi (v) = \phi$ for small
velocity splittings. (Another reason, as we shall see, becomes manifest
when the redshift evolution of clustering is investigated.)

This is certainly the simplest conceivable model of the small-scale
clustering: we have taken everything constant, except for the
absorber number density. Calculation performed using above listed
numerical values of various parameters of the distribution function
and the data set of Cristiani et
al.~(1997) shows that $(\log \xi, \log N_{\rm min})$ curve is very well approximated by a linear dependence giving a value of
\begin{equation}
\label{osam}
\log \phi = 0.16.
\end{equation}
This result is shown in Figure 1. Obvious correlation is accounted
for by our toy model under the assumptions discussed above. The
theoretical slope $a_{\rm th}$ of the linear fit $y=ax+b$ is
determined just
by the index of the column density distribution
\begin{equation}
\label{devet}
a_{\rm th}\equiv \beta-1  =0.55 \pm 0.05,
\end{equation}
(if we consider the best fit of Lu et al.~[1996] as reliable),
independent of $\phi$. We see that the empirical slope
\begin{equation}
\label{deset}
a_{\rm emp}= 0.64 \pm 0.06,
\end{equation}
is equal to the prediction (\ref{devet}) within uncertainties. This fact lends a strong support to our hypothesis.

\begin{figure}
\psfig{file=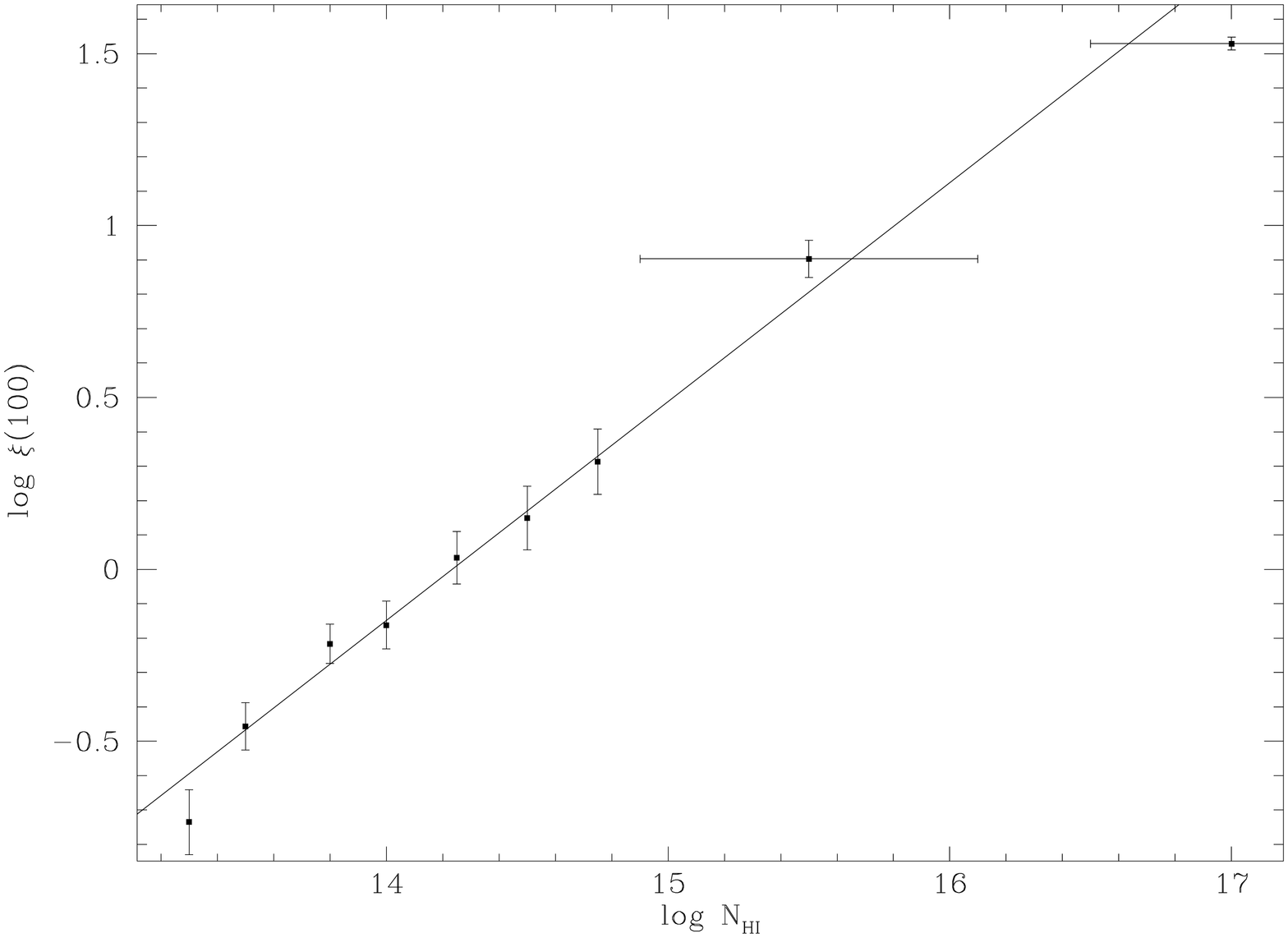,height=6.5cm,angle=270}
\caption{The dependence of the TPCF amplitude (per 100 km s$^{-1}$)
on the column density of ${\rm Ly}\alpha$ and C IV (two highest
column density points) clouds. We see that a linear fit
corresponding to constant clustering on the small scales
is quite satisfactory with the significance of $\sim 84$\%.
It may be noticed that the only significant non-linearity
appears at the smallest column densities, where a diffuse,
truly intergalactic population is expected; the toy model is
inapplicable to these clouds.}
\end{figure}

The value  of $\phi$ in eq.~(\ref{osam}) should be regarded
as the {\it lower limit\/} for small-scale clustering, since it
includes the lowest column density point in Cristiani et al.\
(1996) data, corresponding to the column density below the break
in the distribution (Hu et al.\ 1995), for which not only should
the different value of the exponent in the distribution function
used in evaluating $\phi$, but the very question of the
possibility of the interpretation of these, lowest column density
systems in the framework of our model is doubtful. In Figure 1 we
see that a linear fit corresponding to constant clustering on the
small scales is quite satisfactory with the significance of 84\%.
It may be noticed that the only significant non-linearity
appears at the smallest column densities, where a diffuse, truly intergalactic, population is expected; our simple model does not
apply to these clouds (Fern\'andez-Soto et al.~1996; Gnedin \&
Hui 1996; Weymann et al.~1998). Very low column density Ly$\alpha$
forest is likely to belong to different population of cosmological
objects (Hernquist et al.~1996; Bi \& Davidsen 1997; Rauch 1998).
This is in accordance with findings of Chen et al.~(1998) that
column density of halo clouds sharply declines with galactocentric
distance; since that study also establishes a well-defined maximal
radius for absorption ($R_{\rm max} \sim 174\, h^{-1}$ kpc for
$L_\ast$ galaxies), it is clear that there is a threshold column
density, below which clouds can not be associated with galaxies.
It is obvious that the $\log N_{\rm HI}=13.30$ cm$^{-2}$ point
shows the poorest agreement with the linear fit, and excluding it
from the fit gives unchanged slope $a_{\rm emp}'= 0.59 \pm 0.06$
(showing a satisfactory stability of our model). Although the
value of $\log \xi(\log N=0)'$ is still within uncertainties
equal to the previous value, the central value of $\phi$ is,
however, different by the factor of about 4, since we are dealing
with the unfortunate near-cancellation of two large factors. If
this lowest column density point is discarded, the resulting linear
fit is significant on the $\approx 96$\% level. Although
the number of data points is small, the general
conclusion is that the more realistic fit will tend to give
larger values of $\phi$ and hence the stronger clustering
than that given by eq.~(\ref{osam}).

\begin{figure}
\psfig{file=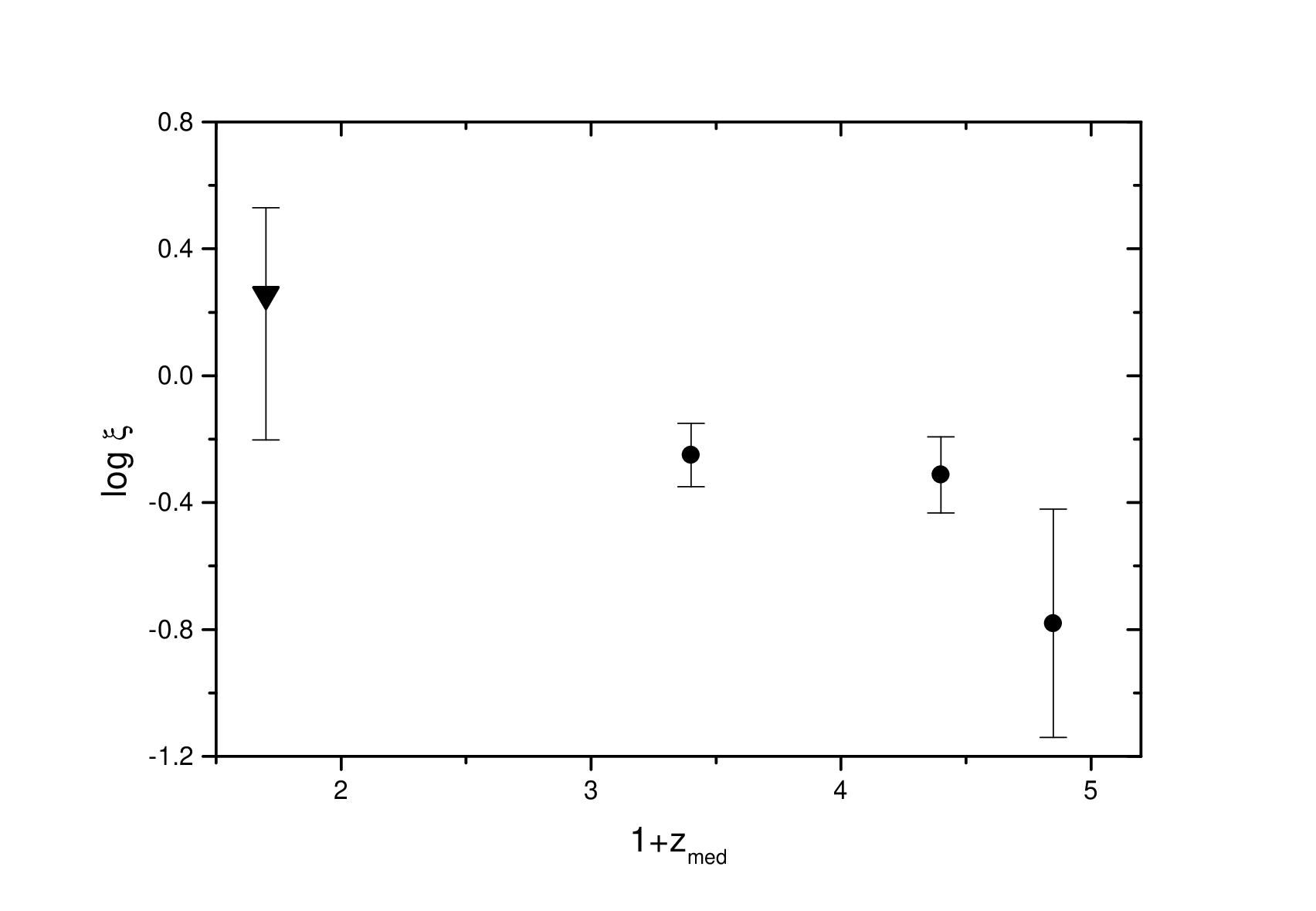,height=6.5cm}
\caption{The redshift dependence of the TPCF amplitude in the
first (100 km s$^{-1}$. With $z_{\rm med}$ we denote the median
redshift in each of the three redshift bins of Cristiani et
al.~(1997). This trend is not noticed at $v>200$ km s$^{-1}$,
which can be accounted for, since only very weak clustering
is expected above some maximum velocity dispersion
$\sigma_{\rm max}$, which value is set by the physics of
extended gaseous haloes.}
\end{figure}

\section{REDSHIFT EVOLUTION}

A trend of decreasing clustering with increasing redshift may also be explained by small-scale constant clustering model. Since the number density of the absorbers counted from any fiducial column density
$N_{\rm min}$ upward increase as
\begin{equation}
\label{dod1}
{{d \cal{N}}\over{dz}} \propto (1+z)^\zeta,
\end{equation}
we could expect from the Equation (\ref{cetiri}) that in a fixed velocity bin, the TPCF amplitude will behave as inverse number density of
absorbing clouds, i.e.
\begin{equation}
\label{jedanes}
\xi \propto (1+z)^{-\zeta - 1},
\end{equation}
i.e.~{\it decrease with increasing
redshift}. Again, this is valid for fixed $v \leq \sigma_{\rm max}$. Although the data presented in Figure 2 are certainly
insufficient to achieve firm conclusions in this regard, they are nevertheless suggestive. We notice
the decrease in the TPCF amplitude quite clearly in the first
($v=100$ km s$^{-1}$) bin, much less pronounced in the second (keep
in mind that we set $\sigma_{\rm max} =150$ km s$^{-1}$), and
completely nonexistent for larger velocity separations. It is very difficult to infer any
quantitative relation from data as such, but we note that the
observed decrease between the first and the third redshift bin
at 100 km s$^{-1}$ separation is within a 20\% from the theoretical
value produced by the simple model, using Kim et al.~(1997) value
for the high-$z$ ${\rm Ly}\alpha$ clouds $\zeta=2.78 \pm 0.71$  (but uncertainties are quite large).  The main conclusion that
clustering {\it decreases\/} with increasing redshift is
incompatible with those classical intergalactic models of  Ly$\alpha$ clouds in which the Hubble expansion and evolution of metagalactic background are only forces driving evolution of absorbing
material, such as pressure-confined models of Sargent et
al.\ (1980) or Ostriker \& Ikeuchi (1983). The data point with
$1+ z_{\rm med} =1.7$ from Ulmer (1996) is included in Fig.~2,
although it corresponds to larger velocity scales and is
not directly comparable to the other data points. Our motivation
here is that it may be regarded as a lower limit for the region of interest; as Ulmer (1996) noted: "...However, the lines appear to be strongly correlated, and the number of expected unresolved pairs
(with $\Delta v < 230$ km s$^{-1}$) is at least as large as the
total number of resolved pairs with 230 km s$^{-1}$ $<\Delta v <
460$ km s$^{-1}$." This is certainly to be expected if clouds are
physically associated with galaxies, since we confidently know that
correlations of the known galactic population were smaller in
the past. Since the large galaxy surveys of our time have become
available, such investigations were performed several times
(Infante \& Pritchet 1992; Bernstein et al.~1994) with clear
result: amplitude of the small-scale clustering increased by
a factor of $\sim 2$ from the $z=0.3$ epoch to the present epoch
($z<0.1$). It should be emphasized that the obstacles in precise quantification of this effect are enormous, as discussed in detail
in the study of Bernstein et al.~(1994). Also, but less significantly,
the theoretical work
on N-body simulations (e.g.~Yoshii, Peterson \& Takahara 1993) came
to the same conclusion about the general trend of the galaxy autocorrelation evolution. Thus, increased clustering of
the Ly$\alpha$ absorbers may
be better understood in framework of the same physical processes which govern the evolution of clustering of normal, luminous galaxies.
The same general trend of increasing clustering with decreasing
redshift is indicated by the data on Ly$\alpha$ forest in the HDF-S (Savaglio et al.~1999).

Additional argument in favor of this simple picture comes from considerations of
influence of the absorbing cloud size on the correlation amplitudes. As
correctly pointed out by Cristiani et al.~(1997), a spatial correlation
function convolved with velocity dispersion produces a correlation
function in the velocity space similar to what is observed if a
cloud sizes
of $\sim 7.5\, h^{-1}$ kpc irrespectively of redshift are assumed.
We propose that this is quite realistic situation, and that realistic velocity dispersions require similar, or even smaller sizes, quite in accordance, for
example, with the sizes obtained by the two-phase gas halo models of Mo (1994) and Mo \& Miralda-Escud\'{e} (1996), Chiba \& Nath (1997) or Miyahata \& Ikeuchi (1995). Such clouds, having total masses
$\sim 10^7 \; M_\odot$ are
similar to progenitors of the present-day globular clusters.
It is indicative that a decrease in the dominant velocity
dispersion scale, which is allowed by all available empirical data (both from autocorrelation
measurements and investigations of close pairs of lines of sight) will result in decrease in sizes of individual contiguous clouds.

\section{DISCUSSION}

This simple picture is what is generally expected from clouds
residing in haloes dominated by dark matter, which are plausible
physical candidates for our points around which absorbers are
clustered with amplitude $\phi (v)$. In physical terms, the dependence
of absorbing column density on distance from the center of an $L_B$ galaxy (Chen et al.~1998)

\begin{eqnarray}
\label{chenmain}
\log \left( \frac{N_{\rm H\ I}}{10^{20} \; {\rm cm}^{-2}}
\right)=  - 5.33 \log \left( \frac{\rho}{10 \; {\rm kpc}} \right) +
    \nonumber\\
     +2.19 \log \left( \frac{L_B}{L_{B_\ast}} \right) + 1.09,&&
\end{eqnarray}
        implies that strong absorption will be seen only near the halo center.
It is straightforward to conclude that these strong and rare absorption
lines have a relatively large probability of having weaker companion
lines originating in the same halo, i.e.~within the galaxy velocity dispersion. Therefore, overall clustering strength is expected to
increase with column density. These absorption sites can be classical
haloes of luminous
galaxies or minihaloes (e.g.~Meiksin 1994; Rauch 1998). Hypothesis
that at least a fraction of Ly$\alpha$ clouds is located in extended
haloes of luminous galaxies has found very strong support in
low-redshift coincident studies (Lanzetta et al.~1996; Chen et al.~1998).
In other words, the results of the present small-scale clustering
analysis presented support the general subclass of models with dark
matter-dominated gravitational confinement. The theoretical basis
of such models is given in important works of Mo (1994) and
Mo \& Miralda-Escud\'e (1996), where the strongly physically motivated
two-phase gaseous halo model has been developed in some detail.
It is not clear at present whether some variant of Black's (1981) classic self-gravitating confinement can also be cast in form which will satisfy the TPCF constraints, but it does not seem very promising. Contrariwise, the theories of origin of the Ly$\alpha$ forest which link the absorption to larger systems, i.e.~clusters or superclusters (Oort 1981; Doroshkevich 1984) are in clear disagreement with these correlation measurements,
due to much higher velocity dispersion of such structures.
The same applies, as correctly noted by Srianand (1996), to the theories involving explosion-type processes
(Ozernoy \& Chernomordik 1978; Chernomordik \& Ozernoy 1983;  Vishniac \& Bust 1987). On the other hand,
strong anticorrelation between low-$z$ Ly$\alpha$ equivalent widths
(i.e.~H I column densities) and galaxy impact parameter in
absorption-selected galaxy sample of Chen et al.\ (1998), indicates that these objects share intragalactic velocity dispersions (i.e.~the same velocity scales as discussed here).
It is difficult, however, to distinguish between models with extended gaseous haloes and huge disks of Maloney (1992). It should be mentioned
that York et al.~(1986) have argued that there are large hydrodynamic
velocities observed in absorption line systems which are similar to those seen in lines of sight through galaxies with active star formation.
Along the same line of thought one should consider the finding of the
{\it HST\/} Key Project (e.g.~Boksenberg 1995) showing that the line density of Ly$\alpha$ absorption systems is greater by nearly an order of
magnitude in the vicinity of metal-line systems (which are believed to originate in haloes of normal galaxies). In the framework of our
toy model, this could be
interpreted as an observational justification for using negligible
$\phi (v)$ outside the range spanned by intragalactic velocities ($0-250$ km s$^{-1}$).
In general, the conclusion that the redshift dependence of the correlation function amplitudes can discriminate between various models of Ly$\alpha$ clouds has important and far-reaching consequences.

Parenthetically, the lack of power in the Ly$\alpha$ absorption line TPCF amplitudes at larger scales in comparison with the TPCF of local
galaxies need not, along the general proposition of Dekel (1982), necessarily be understood as indication for physical difference of
absorber and galaxy populations. Instead, one may follow the suggestion
of Fang (1991) whose results indicate that a biased clustering in the
universe simply has not occurred yet at $z \sim 2$ on large
(comoving) scales. We should keep in mind that absence of large-scale
clustering in the conventional Ly$\alpha$ forest samples has been based mainly on
surveys of high-$z$ absorption lines, usually with $\langle z \rangle \geq
2$. Only large Ly$\alpha$ forest surveys at intermediate and small redshift, likely to be available in the near future, will be capable of definitely solving the puzzle.

We conclude that at this level of accuracy of the TPCF measurements,
a simple model with large and constant small-scale clustering
is able to account for available observational evidence.
Empirically well known galactic velocity dispersion seems to be
capable of entirely explaining observed small-scale clustering
properties of Ly$\alpha$ clouds, in agreement with the Occam's razor.
Much further theoretical work is certainly necessary in order to elaborate the details of the galactic halo model for Ly$\alpha$ clouds. However,
the fact that this model is naturally arising in a compelling
theoretical picture such as the halo cloud model is quite remarkable.

\section*{Acknowledgments}

The authors thank Stefano Cristiani for kindly providing the
necessary data on the TPCF amplitudes, an anonymous referee
for pointing out serious weaknesses in an earlier version
of this manuscript, and Srdjan Samurovi\'c for invaluable
technical help.


\begin{thebibliography}{99}
\bibitem{ } Barcons, X., \& Webb, J. K. 1990, MNRAS, 244, 30p

\bibitem{ } Barcons, X., \& Webb, J. K. 1991, MNRAS, 253, 207

\bibitem{ } Bernstein, G. M., Tyson, J. A., Brown, W. R., \& Jarvis, J. F. 1994, ApJ, 426, 516

\bibitem{ } Binney, J., \& Tremaine, S. 1987, Galactic dynamics
(Princeton: Princeton University Press)

\bibitem{ } Black,  J. H. 1981, MNRAS, 197, 553

\bibitem{ } Boksenberg, A. 1995, in QSO Absorption Lines, ed.~by G. Meylan (Berlin: Springer), 253

\bibitem{ } Carbone, V., \& Savaglio, S. 1996, MNRAS, 282, 868

\bibitem{ } Cen, R., \& Simcoe, R. A 1997, ApJ, 483, 8

\bibitem{ } Chen, H.-W., Lanzetta, K. M., Webb, J. K., \& Barcons, X.
1998, ApJ, 498, 77

\bibitem{ } Chernomordik, V. V. 1995, ApJ, 440, 431

\bibitem{ } Chernomordik, V. V., \& Ozernoy, L. M. 1983, Nature 303, 153

\bibitem{ } Chiba, M., \& Nath, B. B. 1997, ApJ, 483, 638

\bibitem{ } Cowie, L. L., Songaila, A., Kim, T.-S., \& Hu, E. M. 1995, AJ, 109, 1522

\bibitem{ } Cristiani, S., D'Odorico, S., D'Odorico, V., Fontana, A., Giallongo, E., \& Savaglio, S. 1997, MNRAS, 285, 209

\bibitem{ } Crotts, A. P. S. 1989, ApJ, 336, 550

\bibitem{ } Dekel, A. 1982, ApJ, 261, L13

\bibitem{ } Doroshkevich, A. G. 1984, AZh, 61, 218

\bibitem{ } Elowitz, R. M., Green, R. F., \& Impey, C. D. 1995, ApJ, 440, 458

\bibitem{ } Fang, L. Z. 1991, A \& A, 244, 1

\bibitem{ } Fardal, M. A., \& Shull, J. M. 1993, ApJ, 415, 524

\bibitem{ } Fern\'andez-Soto, A., Lanzetta, K. M., Barcons, X., Carswell, R. F., Webb, J. K., \& Yahil, A. 1996, ApJ, 460, L85

\bibitem{ } Gnedin, N. Y., \& Hui, L. 1996, ApJ, 472, L73

\bibitem{ } Heisler, J., Hogan, C. J., \& White, S. D. M. 1989, ApJ, 347, 52

\bibitem{ } Hu, E. M., Kim, T.-S., Cowie, L. L., \& Songaila, A. 1995, AJ, 110, 1526

\bibitem{ } Infante, L., \& Pritchet, C. J. 1992, ApJS, 83, 237

\bibitem{ } Jannuzi, B. T. 1997, in Structure and Evolution of the Intergalactic Medium from QSO Absorption Line Systems, ed.~P. Petitjean \& S. Charlot (Paris: Edition Fronti\'eres), 93

\bibitem{ } Jannuzi, B. T., et al. 1998, ApJS, 118, 1

\bibitem{ } Kim, T.-S., Hu, E. M., Cowie, L. L., \& Songaila, A. 1997, AJ, 114, 1

\bibitem{ } Lanzetta, K. M., Webb, J. K., \& Barcons, X. 1996, ApJ,
456, L17

\bibitem{ } Liu, X. D., \& Jones, B. J. T. 1990, MNRAS, 242, 678

\bibitem{ } Lu, L., Sargent, W. L. W., Womble, D. S., \& Masahide, T.-H. 1996,
ApJ, 472, 509

\bibitem{ } Maloney, P. 1992, ApJ, 398, L89

\bibitem{ } Meiksin, A. 1994, ApJ, 431, 109

\bibitem{ } Meiksin, A., \& Bouchet, F. R. 1995, ApJ, 448, L85

\bibitem{ } Miyahata, K., \& Ikeuchi, S. 1995, PASJ, 47, L37

\bibitem{ } Mo, H. J. 1994, MNRAS, 269, L49

\bibitem{ } Mo, H. J., \& Miralda-Escud\'{e}, J. 1996, ApJ, 469, 589

\bibitem{ } Mo, H. J., Xia, X. Y., Deng, Z. G., Boerner, G., \& Fang, L. Z. 1992, A \& A, 256, L23

\bibitem{ } Muecket, J. P., \& Mueller, V. 1987, Ap \& SS, 139, 163

\bibitem{ } Oort, H. J. 1981, A \& A, 94, 359

\bibitem{ } Ostriker, J. P., Bajtlik, S., \& Duncan, R. C. 1988, ApJ, 327, L35

\bibitem{ } Ozernoy, L. M., \& Chernomordik, V. V. 1978, Sov. Astron. 22, 141

\bibitem{ } Pando, J., \& Fang, L.-Z. 1996, ApJ, 459, 1

\bibitem{ } Rauch, M., Carswell, R. F., Chaffee, F. H., Foltz, C. B., Webb, J. K., Weymann, R. J., Bechtold, J., \& Green, R. F. 1992, ApJ, 390, 387

\bibitem{ } Rauch, M. 1998, ARAA, 36, 267

\bibitem{ } Salmon, J., \& Hogan, C. 1986, MNRAS, 221, 93

\bibitem{ } Sargent, W. L. W., Young, P. J., Boksenberg, A., \& Tytler, D. 1980, ApJS, 42, 41

\bibitem{ } Savaglio, S. et al.\ 1999, ApJ, 515, L5

\bibitem{ } Srianand, R. 1996, ApJ, 462, 68

\bibitem{ } Srianand, R., \& Khare, P. 1994, MNRAS, 271, 81

\bibitem{ } Ulmer, A. 1996, ApJ, 473, 110

\bibitem{ } Vishniac, E. T., \& Bust, G. S. 1987, ApJ, 319, 14

\bibitem{ } Webb, J. K. 1987, in Observational Cosmology, ed. A. Hewitt, G. Burbidge, \& L. Z. Fang (Dordrecht: Reidel/Kluwer), p. 803

\bibitem{ } Webb, J. K., \& Barcons, X., 1991, MNRAS, 250, 270

\bibitem{ } Weymann, R. J., et al. 1998, ApJ, 506, 1

\bibitem{ } White, S. D. M., \& Rees, M. J. 1978, MNRAS, 183, 341

\bibitem{ } York, D. G., Dopita, M., Green, R. \& Bechtold, J. 1986, ApJ,
311, 610

\bibitem{ } Yoshii, Y., Peterson, B. A., \& Takahara, F. 1993, ApJ, 414, 431
\end{thebibliography}
\end{document}